\documentclass[twocolumn,showpacs,preprintnumbers,amsmath,amssymb,prb]{revtex4}
\usepackage{graphicx}
\usepackage{dcolumn}
\usepackage{bm}

\begin{document}
\title{Disordered two-dimensional superconductors: \\
roles of temperature and interaction strength} 

\author{Felipe \surname{Mondaini},$^1$  
        Thereza \surname{Paiva},$^1$ 
        Raimundo R. \surname{dos Santos},$^1$
        and 
        R.\ T.\ \surname{Scalettar},$^2$} 

\affiliation{$^{1}$Instituto de F\'\i sica, 
                 Universidade Federal do Rio de Janeiro, 
                 Caixa Postal 68528,
                 21941-972 Rio de Janeiro RJ, 
                 Brazil\\
             $^2$Physics Department,
                 University of California
                 Davis CA 95616, 
                 USA}

\begin{abstract}
We have considered the half-filled disordered attractive Hubbard model on a square lattice, in which the on-site attraction is switched off on a fraction $f$ of sites, while keeping a finite $U$ on the remaining ones. 
Through Quantum Monte Carlo (QMC) simulations for several values of $f$ and $U$, and for system sizes ranging from $8\times 8$ to $16\times 16$, we have calculated the configurational averages of the equal-time pair structure factor $P_s$, and, for a more restricted set of variables, the helicity modulus, $\rho_s$, as functions of temperature.  
Two finite-size scaling {\it ansatze} for $P_s$ have been used, one for zero-temperature and the other for finite temperatures. 
We have found that the system sustains superconductivity in the ground state up to a critical impurity concentration, $f_c$, which increases with $U$, at least up to $U=4$ (in units of the hopping energy). 
Also, the normalized zero-temperature gap as a function of $f$ shows a maximum near $f\sim 0.07$, for $2\lesssim U\lesssim 6$. 
Analyses of the helicity modulus and of the pair structure factor led to the determination of the critical temperature as a function of $f$, for $U=3,$ 4 and 6: they also show maxima near $f\sim 0.07$, with the highest $T_c$ increasing with $U$ in this range.
We argue that, overall, the observed behavior results from both the breakdown of CDW-superconductivity degeneracy and the fact that free sites tend to ``push" electrons towards attractive sites, the latter effect being more drastic at weak couplings.    

\end{abstract} 

\date{\today}


\maketitle

\section{Introduction}
\label{intro}
The interplay between impurities and superconducting pairing has been a challenging problem for some time.\cite{Goldman98} 
It has been argued \cite{Anderson59} that as long as electronic states remain extended in the presence of weak disorder, superconductivity should not be affected; as disorder increases, however, superconductivity must eventually be suppressed. 
The consequences of this competition are especially interesting in two dimensions, since the superconducting transition belongs to the $xy$-model universality class (the Kosterlitz-Thouless transition to a state without long range order), while non-interacting electrons become localized in the presence of any amount of disorder. 
Indeed, by varying the thickness of thin films of Bi and Pb deposited on Ge substrates,\cite{Haviland89} the sheet resistance $R_\square$ shows insulating behavior (i.e., it increases with decreasing temperature $T$) for the thinner samples, and superconducting behavior for the thicker samples; the separatrix between these two regimes extrapolates to a quantum critical point as $T\to 0$,\cite{Goldman98} and one of the unresolved issues is whether or not the threshold of $R_\square$, $R^*$, is universal, i.e., $R^*=R_Q\equiv h/4e^2$.

From the theoretical point of view, the two-dimensional behavior has been examined in a variety of ways. Some have exploited a bosonic description of Cooper pairs,\cite{Fisher89,Fisher90} according to which electrons near the Fermi surface are paired and localization is driven by Coulomb repulsion amongst pairs. On the other hand, one may envisage a fermionic mechanism: disorder enhances Coulomb repulsion amongst electrons, thus decreasing the effective screening (due to electron-phonon interactions, in conventional superconductors) which in turn leads to the disappearance of Cooper pairs.\cite{Larkin99} Intermediate pictures have also been proposed\cite{Feigelman01} within a phenomenological theory to address the issue of universality of $R^*$. 

Since none of these approaches have succeeded in fully explaining experimental data, alternative routes should be sought. One possibility is to study simplified microscopic fermionic models in which disorder is incorporated in a fundamental and unbiased way. However, not much is known about models in which disorder is present in the pairing interaction. With the purpose of bridging this gap, and due to the fact that disorder is more readily dealt with in real-space, here we consider the disordered attractive Hubbard model, whose Hamiltonian reads, 
\begin{eqnarray}
\label{Ham}
H &=& -t\sum_{\langle {\bf rr'} \rangle \sigma}
\left( c_{{\bf r} \sigma}^{\dagger}c_{{\bf r'}\sigma}^{\phantom{\dagger}} 
+ \text{H.c.}\right) 
-\mu \sum_{{\bf r} \sigma} n_{{\bf r} \sigma}
\nonumber\\
&-& \sum_{{\bf r}} \, U({\bf r}) \, n_{{\bf r} \uparrow}
                    n_{{\bf r} \downarrow},
\end{eqnarray}
where $c^{\dagger}_{{\bf r}\sigma} 
(c_{{\bf r}\sigma}^{\phantom{\dagger}})$ 
are fermion creation (destruction) operators at site ${\bf r}$
with spin $\sigma$, and $n_{{\bf r}\sigma}=
c_{{\bf r}\sigma}^{\dagger}c_{{\bf r}\sigma}^{\phantom{\dagger}}$, and 
H.c. stands for Hermitian conjugate of the previous term.
The kinetic energy lattice sum $\langle {\bf rr'} \rangle$ is over
nearest-neighbor sites on a two-dimensional square lattice,
and $\mu$ is the chemical potential; the hopping integral sets the energy scale, 
so we take $t=1$ throughout this paper.  
The on-site attraction $U({\bf r})$ is chosen to take on the two values
$U({\bf r})=0$ and $U({\bf r})=U$ with probabilities $f$ and $1-f$ respectively; 
note that $U>0$ corresponds to attraction, according to our definition of the on-site term in (\ref{Ham}).
A review of the homogeneous model can be found in Ref.\ \onlinecite{Micnas90}, while Ref.\ \onlinecite{Paiva03} deals with a recent extension of the model to describe non-random layered superconductors, as the borocarbides. 

The above model mimics the thin films of Bi and Pb referred to above, in the sense that the inverse film thickness tracks the concentration, $1-f$, of attractive sites.\cite{Haviland89} Additionally, it also describes the effects of negative-$U$ centers, which are thought to be relevant to high-temperature superconductivity in the cuprates.\cite{Micnas90,Wilson01} This model has been studied at mean-field level,\cite{Litak00,Aryanpour06,Aryanpour07} and the main results, for a given electronic density, can be summarized as follows: (i) superconductivity in the ground state is destroyed for impurity concentrations above $f_c$; (ii) $f_c$ decreases as $U$ increases [very simple heuristic arguments\cite{Litak00} lead to $f_c= 1-(U/W)^2$ in two dimensions, with $W=8t$ being the bandwidth]; and (iii) $T_c(f)/T_c(0)$ is a concave function of $f$, which vanishes at $f_c(U)$. Though mean-field approximations are useful as a first approach to the problem, one should be extremely cautious about their predictions for two dimensional systems. For instance, for the pure system at half filling, the degeneracy of charge-density wave and superconducting order leads to an effective three-component order parameter, thus suppressing the critical temperature to zero\cite{Scalettar89,Moreo91,Paiva04} by virtue of the Mermin-Wagner theorem;\cite{Mermin66} mean-field approaches are unable to detect this feature, and should therefore lead to unreliable results close to half filling.
Indeed, recent Quantum Monte Carlo (QMC) simulations have predicted that a small amount of disorder at half filling initially enhances superconductivity;\cite{Hurt05} this was attributed to the impurity-induced breakdown of the above-mentioned degeneracy. In view of all this, a more thorough investigation of the model at half filling is clearly in order. 
Here we report on Quantum Monte Carlo studies of the dependence of $f_c$ with $U$ as well as of the dependence of $T_c$ with $f$, for different values of $U$; we recall that Ref.\ \onlinecite{Hurt05} was restricted to $U=4$ and $T=0$ only. As we will see, our predictions are very different from those of mean-field approaches. 

The paper is organized as follows. In Sec.\ \ref{QMC} we outline the QMC method, and discuss the quantities used to locate the superconducting transitions. In Sec.\ \ref{gs} we present a finite-size scaling (FSS) analysis of data for superconducting correlations in the ground state, from which we extract the behavior of $f_c$ with $U$. 
In Sec.\ \ref{finiteT} we perform finite-temperature FSS analyses of data for superconducting correlations, as well as analyses of the superfluid density, to obtain $T_c(f)$ for different values of $U$. 
And, finally, section \ref{concs} summarizes our findings.

\section{The Computational Approach}
\label{QMC}

We use the determinant QMC method \cite{Blankenbecler81,Hirsch83c,Hirsch85,White89,dosSantos03b} 
to investigate the ground-state as well as finite-temperature properties of the model. 
In this approach, the imaginary-time interval $(0; \beta)$ is discretized into $M$
slices separated by the interval $\Delta\tau$ and a path integral expression is
written down for the partition function $Z$. The electron-electron
interactions are decoupled by the introduction of a 
Hubbard-Stratonovich field.\cite{Hirsch83c} The fermion
degrees of freedom can then be integrated out analytically,
leaving an expression for $Z$ which involves an integral over
the Hubbard-Stratonovich field, 
with an integrand which is the product 
of two determinants of matrices of dimension the system size.
We perform the integral stochastically.  
In the case of the attractive Hubbard model
considered here, the traces over the spin up and spin down electrons are
given by the determinant of the same matrix,
the integrand is a perfect square,
and hence there is no sign problem.\cite{Hirsch85,dosSantos03b}

In order to study the physics at a particular lattice size $L\times L$
and value of $f$,
we randomly choose $fL^2$ sites and set $U=0$ on those sites.  
We typically use 30--50 
such realizations to average over the different
disorder configurations.
If $fL^2$ is not an integer, we average over the two adjacent
integer values, with appropriate weights.    
For each disorder configuration, observables are evaluated
as the appropriate combinations of Green's functions, which are given
by matrix elements of the inverse of the matrix appearing
as the integrand.\cite{Blankenbecler81,Hirsch85,White89,dosSantos03b}
The average over different disorder configurations then yields the quantities of interest. 
Systematic errors in the calculated quantities, 
associated with our choice of $\Delta \tau$ for the discretization of $\beta$,
are typically smaller than both the error bars associated with the statistical fluctuations
for a single disorder realization, and the error bars associated
with sample-to-sample variations. 

As discussed previously,\cite{Hurt05} a useful quantity to locate the transition is the 
configurationally-averaged equal-time pairing structure factor,
\begin{equation}
P_s = \left[\sum_{{\bf r}} \Gamma ({\bf r})
\right],
\label{Ps}
\end{equation} 
where $[\ldots]$ denotes average over disorder configurations (thus restoring translational invariance), and the pairing correlation function is
\begin{equation}
\Gamma({\bf r})\equiv \langle \Delta({\bf i})
\Delta^\dagger
({\bf i}+{\bf r})+ {\rm H.c.}\rangle\ ,
\label{Gr}
\end{equation}
where $\langle\ldots\rangle$ denotes ensemble average, with
\begin{equation}
\Delta ({\bf r}) 
= c_{{\bf r}\downarrow} c_{{\bf r}\uparrow}.
\label{Delta}
\end{equation}

The different scaling behaviors of $P_s$, in the ground state and at finite temperatures, will be discussed in subsequent sections. 

Further, current-current correlations probe the superfluid weight
and provide an alternative way to detect the
destruction of superconductivity.\cite{Scalapino93}  
We define
\begin{equation}
{\Lambda}_{xx}({\bf r}, \tau)
= \langle j_{x} ({\bf r}, \tau) j_{x} (0, 0) \rangle\;,
\label{lambda}
\end{equation}
where
\begin{equation}
j_{x}({\bf r}, \tau) = e^{H \tau}
\left[it \sum_\sigma
(c_{{\bf r}+\hat x,\sigma}^{\dagger}c_{{\bf r},\sigma}^{\ } -
c_{{\bf r},\sigma}^{\dagger} c_{{\bf r}+\hat x,\sigma}^{\ } )
\right] e^{-H \tau}\;,
\label{jx}
\end{equation}
and the Fourier transform in space and imaginary time,
\begin{equation}
\Lambda_{xx}({\bf q},\omega_n) = \frac{1}{N_s}\sum_{\bf r} \int_0^\beta d \tau
e^{i {\bf q}\cdot{\bf r} }
e^{-i \omega_n \tau}
\Lambda_{xx}({\bf r},\tau),
\end{equation}
where $N_s$ is the number of lattice sites, and $\omega_n = 2n\pi/\beta$.

The longitudinal part of the current-current correlation function
satisfies the f-sum rule, which relates its value to the kinetic energy
$K_x$, 
\begin{equation}
\Lambda^{{\rm L}}  \equiv 
{\rm lim}_{q_x \rightarrow 0} 
{\Lambda}_{xx} (q_{x},q_{y}=0,\omega_n = 0)  
\end{equation}
\begin{equation}
\Lambda^{\rm L} = K_{x}\;,
\end{equation}
where $K_{x} = \langle -t \sum_\sigma
(c_{{\bf r}+\hat x,\sigma}^{\dagger}c_{{\bf r},\sigma}^{\ } +
c_{{\bf r},\sigma}^{\dagger} c_{{\bf r}+\hat x,\sigma}^{\ } ) \rangle$.
Meanwhile, in the superconducting state the transverse part,
\begin{equation}
\Lambda^{\rm T}\equiv
{\rm lim}_{q_y \rightarrow 0} \hskip0.1in
{\Lambda}_{xx} (q_{x}=0,q_{y},\omega_n=0) \;,
\end{equation}
can differ from the longitudinal part,
the difference being
the superfluid stiffness $D_s$,
\begin{equation}
D_{s}/\pi  =  [
\Lambda^{\rm L}-
\Lambda^{\rm T}]
 =  
[K_{x}-
\Lambda^{\rm T}]\;.
\label{stiff}
\end{equation}
Thus the current-current correlations provide an alternative, 
complementary method to the equal time
pair correlations for looking at the superconducting transition.

\begin{figure}[t]
{\centering\resizebox*{8.5cm}{!}{\includegraphics*{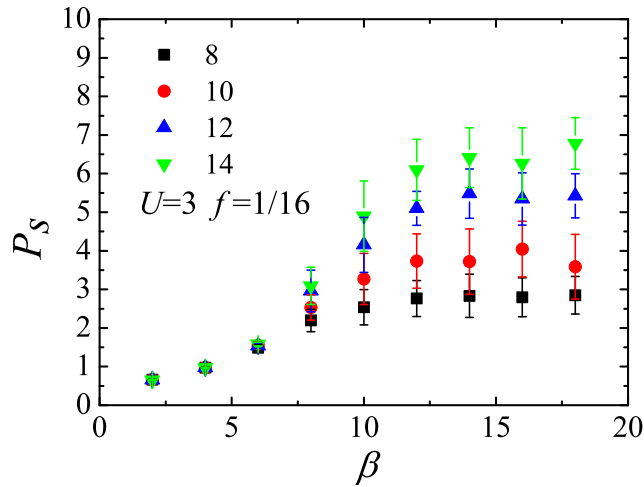}}}
\caption{(Color online) Configurationally-averaged equal-time pair structure factor $P_s$ as a function of the inverse temperature $\beta$, for different lattice sizes at half filling and concentration of free sites $f=1/16$.}
\label{fig1} 
\end{figure}

\section{Ground state properties: $f_c(U)$}
\label{gs}

As remarked above, there is no `sign problem' for the attractive Hubbard
model, so we can do computations at very low temperatures 
(large $\beta$), as shown in Fig.\ \ref{fig1} for the data of unscaled $P_s$. 
We recall that throughout this paper we only consider the case of a half-filled band.
The error bars result from the dispersion in the average values taken over disorder configurations. 

The finite-size scaling behavior of $P_s$ allows us to extract quantitative 
information about the superconducting transition. 
As shown by Huse,\cite{Huse88} the spin-wave
correction to the pair structure factor in the ground state is expected to be inversely proportional
to the linear lattice size,
\begin{equation}
\frac{P_s}{L^2}=|\Delta_0|^2+ \frac{a}{L},
\label{Huse}
\end{equation}
where $\Delta_0$ is the superconducting gap function at zero temperature, 
and $a\equiv a(U,f)$ is independent of $L$.

\begin{figure}[h]
{\centering\resizebox*{8.2cm}{!}{\includegraphics*{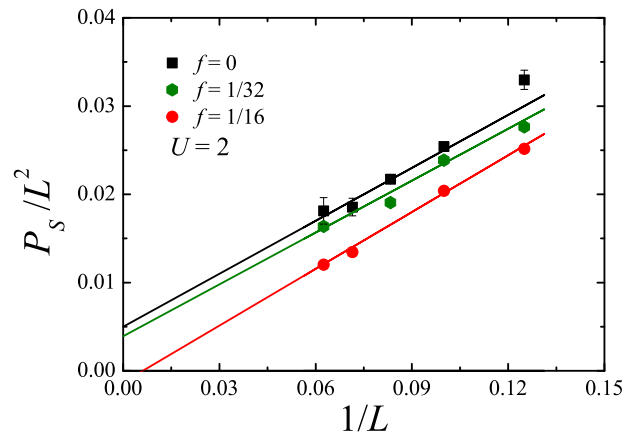}}}
\caption{(Color online) Zero-temperature scaling of the 
configurationally-averaged equal-time pair structure factor, for $U=2$, 
and for different disorder concentrations, $f$. For each $f$, the intersection 
with the vertical axis yields the (squared) zero-temperature gap; see Eq.\ (\ref{Huse}).}
\label{PsLU2} 
\end{figure}
\begin{figure}
\bigskip
{\centering\resizebox*{8.2cm}{!}{\includegraphics*{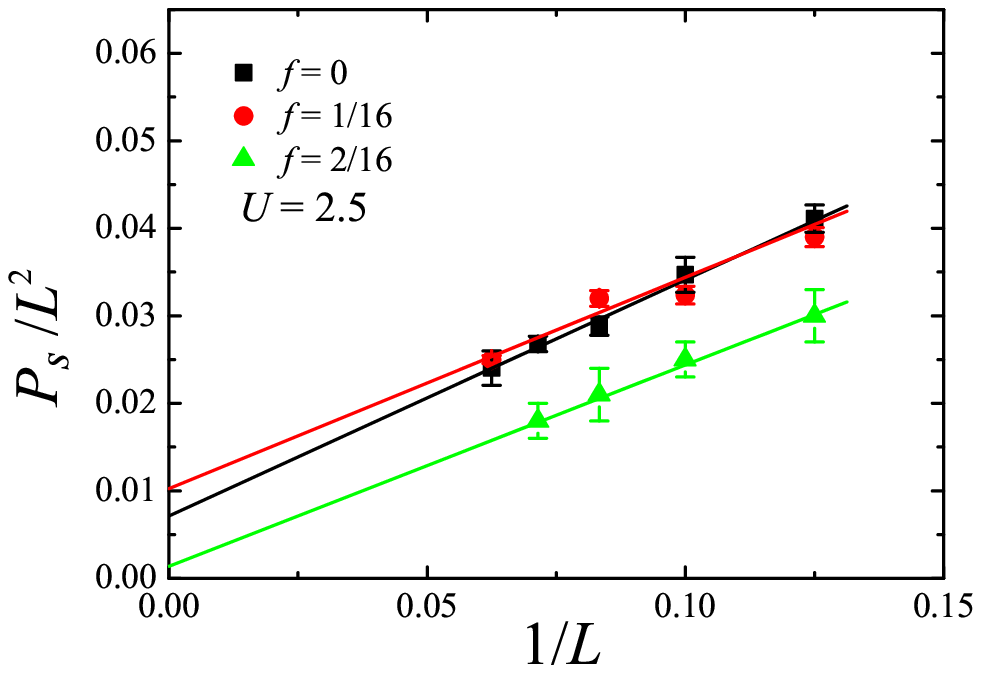}}}
\caption{(Color online) Same as Fig.\ \ref{PsLU2}, but for $U=2.5$.}
\label{PsLU2.5} 
\end{figure}
\begin{figure}
\bigskip
{\centering\resizebox*{8.2cm}{!}{\includegraphics*{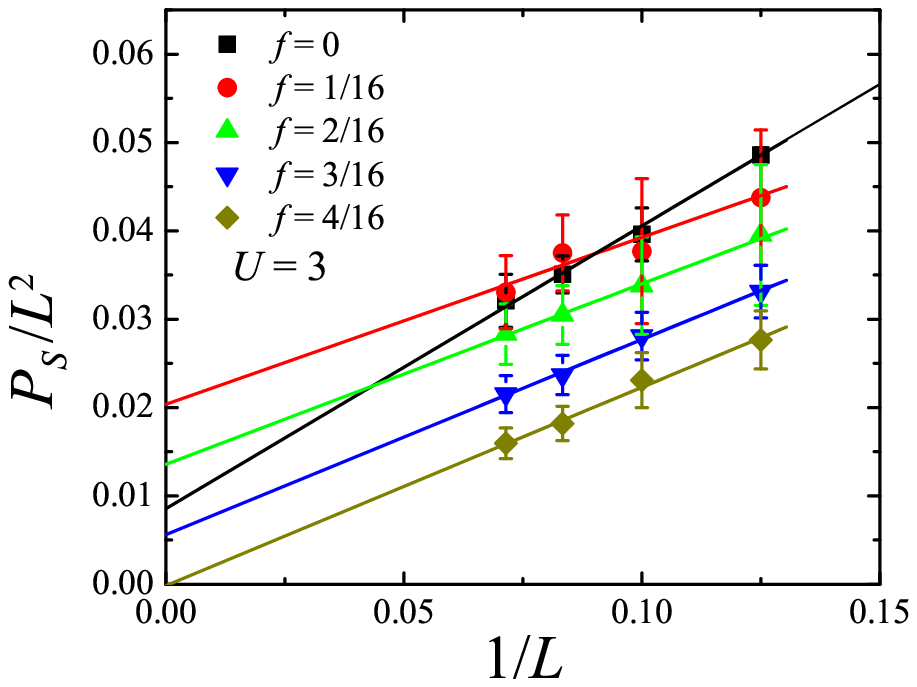}}}
\caption{(Color online) Same as Fig.\ \ref{PsLU2}, but for $U=3$.}
\label{PsLU3} 
\end{figure}
\begin{figure}
{\centering\resizebox*{8.2cm}{!}{\includegraphics*{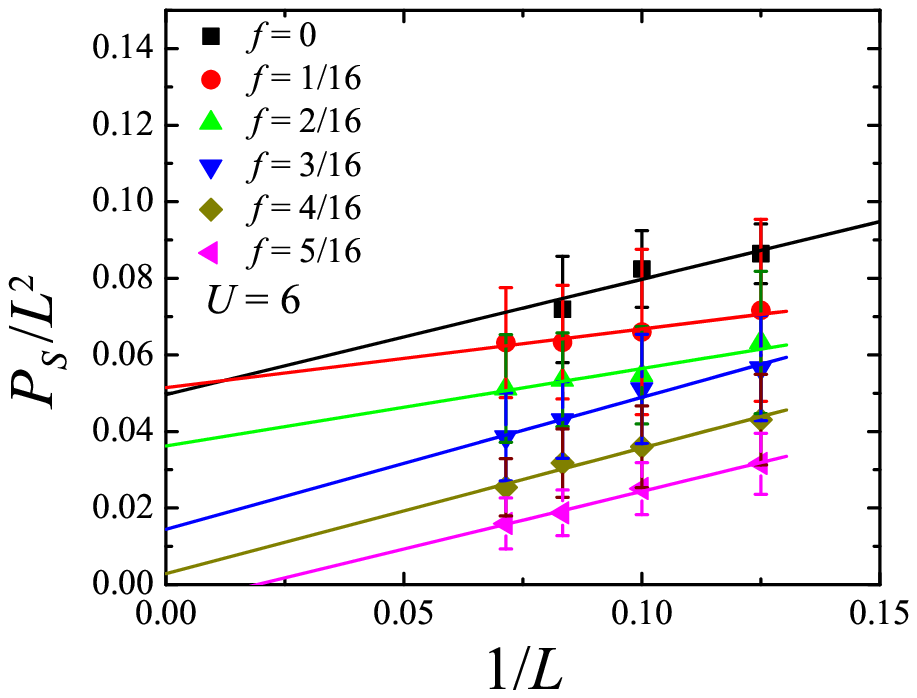}}}
\caption{(Color online) Same as Fig.\ \ref{PsLU2}, but for $U=6$.}
\label{PsLU6} 
\end{figure}

\begin{figure}
{\centering\resizebox*{8.5cm}{!}{\includegraphics*{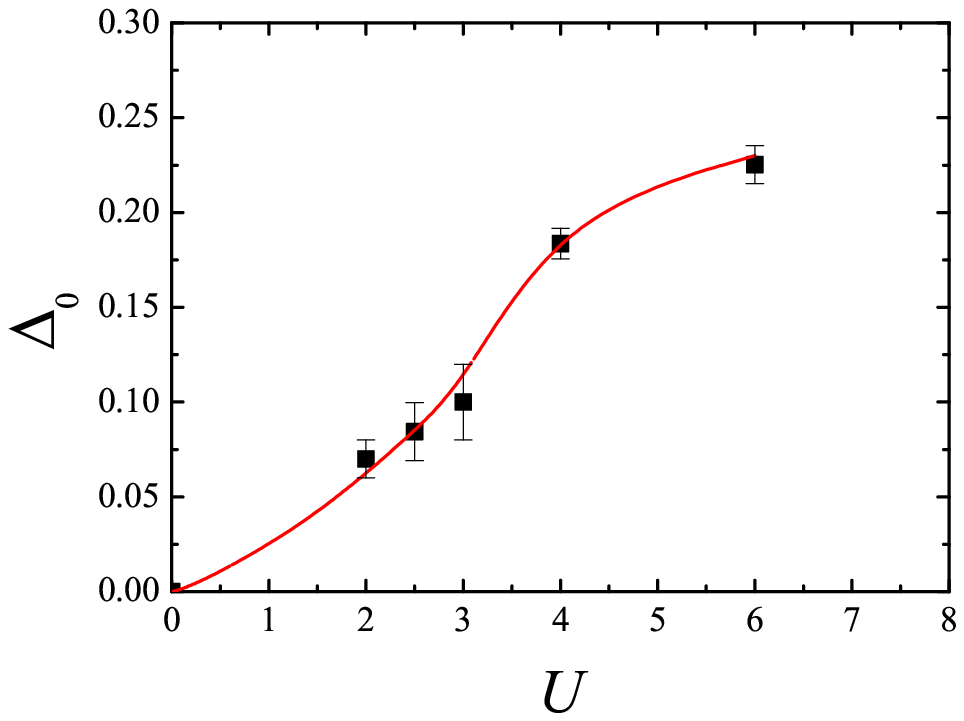}}}
\caption{(Color online) Zero-temperature gap for the pure system, obtained from the $f=0$ data, 
together with Eq.\ (\ref{Huse}). The full line is a guide to the eye.}
\label{Delta0} 
{\centering\resizebox*{8.5cm}{!}{\includegraphics*{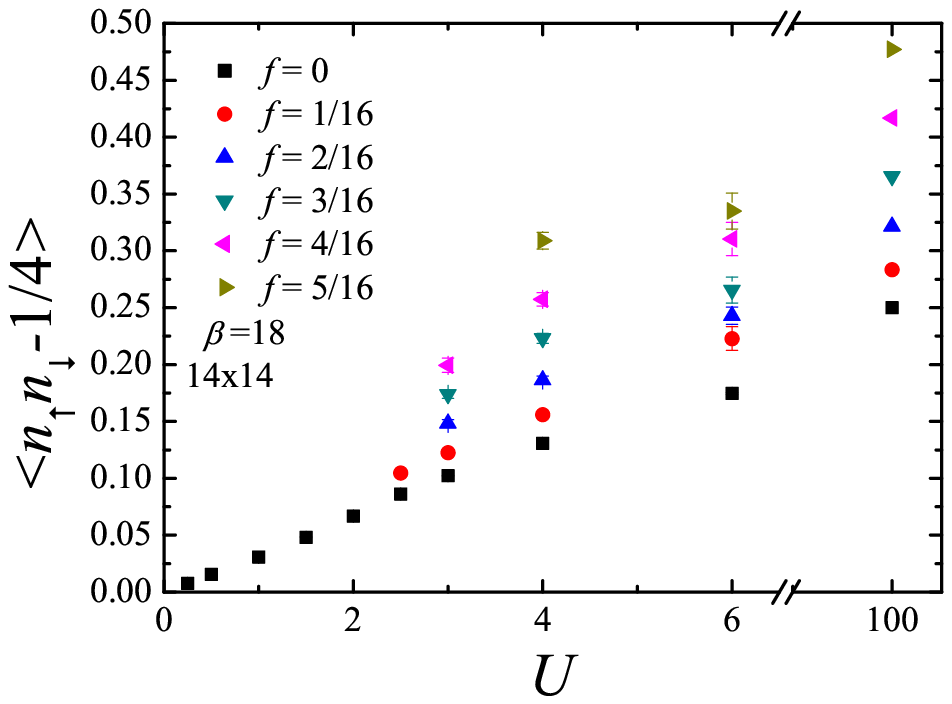}}}
\caption{(Color online) Average double occupancy on attractive sites as a function of $U$, for the pure case and for several disordered configurations. Data are for $14\times 14$ lattices and for $\beta=18$. Data for $U=100$ correspond to the strong coupling limit, Eq.\ \ref{double}.}
\label{nupndn} 
\end{figure}

In Figs.~\ref{PsLU2}--\ref{PsLU6} we plot the $T\to 0$ extrapolated values of
$P_s/L^2$ versus $1/L$, for linear lattice sizes ranging from
$L=8$ to $L=16$, and $U=2$, 2.5, 3 and 6 (data for $U=4$ can be found in 
Fig.\ 4 of Ref.\ \onlinecite{Hurt05}). 
According to Eq.\ (\ref{Huse}), each intercept with the vertical axis 
provides an estimate for $\Delta_0^2$ (the square of the zero-temperature gap) for the values of $U$ and $f$ considered.
For the pure system ($f=0$), $\Delta_0$ is plotted as a function of $U$ in 
Fig.\ \ref{Delta0}: the observed increase of $\Delta_0$ (at least up to $U\sim 6$), is due to an increase in the average site double occupancy. 
Indeed, Fig.\ \ref{nupndn} shows the double occupancy on attractive sites, $d_A\equiv\langle n_\uparrow n_\downarrow - 1/4\rangle$, as a function of $U$ for both the pure system and for several disordered configurations. 
The overall behavior is an increase in $d_A$ with $U$; and, for a fixed $U$, this double occupancy increases with $f$, as it can be seen from its strong coupling limit, 
\begin{equation}
\langle n_\uparrow n_\downarrow - 1/4\rangle=\frac{1}{4}\frac{1+f}{1-f},\ {\rm for}\ U\to\infty.
\label{double}
\end{equation}

\begin{figure}
{\centering\resizebox*{8.5cm}{!}{\includegraphics*{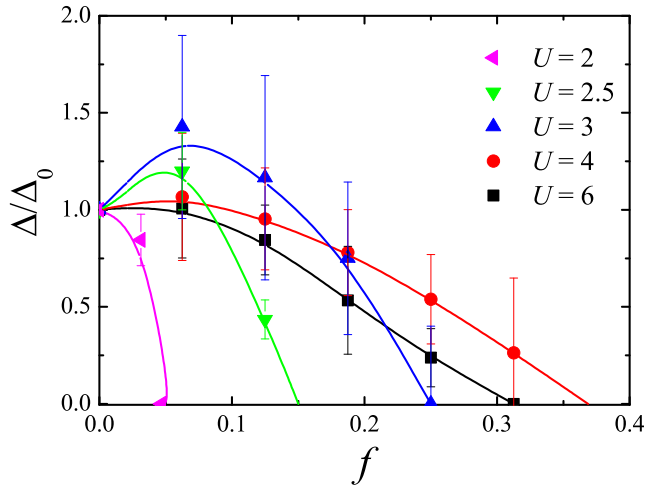}}}
\caption{(Color online) Normalized zero-temperature gaps as functions of impurity concentration, for different values of $U$. Full lines are guides to the eye.}
\label{gaps} 
\end{figure}
  
In order to compare the effects of disorder for different attraction intensities, for each $U$ we normalize the zero-temperature gap, $\Delta$, by their respective pure system values, $\Delta_0$; the result is displayed in Fig.\ \ref{gaps}.   
For $2.5\lesssim U < 6$, the normalized gaps initially increase with 
disorder, reaching maxima $\left[\Delta/\Delta_0\right]_{\rm max}$ around $f\sim 0.07$. 
It should be noticed that $\left[\Delta/\Delta_0\right]_{\rm max}$, in turn, does not behave monotonically as a function of $U$, but displays a maximum for $U=3$, amongst the values of $U$ examined.
Another crucial information extracted from Fig.\ \ref{gaps} is that a maximum of $\Delta/\Delta_0$ is absent for $U=6$. 
This different behavior for larger $U$ therefore indicates that the disorder-induced breakdown of 
CDW-superconductivity degeneracy is not the only mechanism at play: for smaller $U$, the presence of free sites contributes to a decrease in the single occupancy, by `pushing' electrons to the attractive sites. For larger $U$, the pairs are so tightly bound that the relative weight of single occupancy is smaller, and disorder has hardly any effect on forcing the electrons to occupy the attractive sites. 
Figure \ref{nupndn} indeed shows that the percentual enhancement in double occupancy due to disorder is  larger for $U\sim 4$ than for $U\sim 6$. 
As disorder increases, the presence of free sites strongly disturbs pair coherence, and the gap decreases. 
The initial increase with disorder has also been predicted for anisotropic superconductors with mesoscopic phase separation.\cite{Coleman95,Yukalov04}

\begin{figure}
{\centering\resizebox*{8.5cm}{!}{\includegraphics*{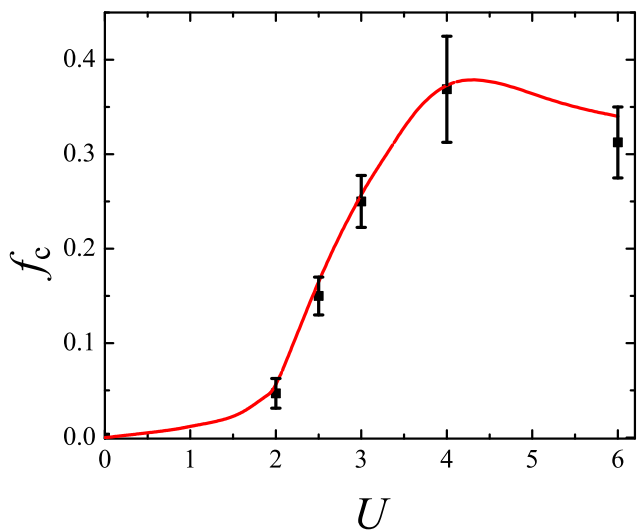}}}
\caption{(Color online) Critical impurity concentration as a function of $U$, obtained from Fig.\ \ref{gaps}. The full line is a guide to the eye.}
\label{fcU} 
\end{figure}

We can obtain the dependence of $f_c$ with $U$ by extrapolating the data for 
$\Delta/\Delta_0$ to zero. The intercept with the 
horizontal axis of each of the curves in Fig.~\ref{gaps} provides $f_c$ for the 
corresponding $U$, and the result is displayed in Fig.\ \ref{fcU}; 
the error bars reflect the uncertainties in the extrapolations of $\Delta/\Delta_0$ to 
zero in Fig.\ \ref{gaps}. 
It should be noted that $f_c$ initially (i.e., for $U\lesssim 2$) increases very slowly with $U$, which should be attributed to the fact that the pairs are not so strongly bound for small $U$, so that a small amount of free sites destroys phase coherence.  
As $U$ increases, the pairs become more tightly bound, and a larger amount of disorder can be sustained before the free sites switch their roles, from pushing electrons onto attractive sites to that of destroying phase coherence. 
This behavior is in disagreement with mean-field predictions, according to which $f_c$ should decrease with $U$.\cite{Litak00} However, this switching of roles played by the free sites is an effect too subtle to be picked up by approaches, such as mean-field ones, which do not incorporate fluctuations in a fundamental way.  
For $U\gtrsim 4$, $f_c$ appears to be decreasing with $U$, and the agreement with the mean-field approach would set in.

We recall that the strong-coupling 
pure attractive Hubbard model (at half filling) can be mapped onto an 
isotropic Heisenberg model.\cite{Emery76,dosSantos93}    
Nonetheless, the fact that in the disordered case, $f_c$ displays such strong dependence with $U$ is a 
clear indication that the mechanisms of superconductivity suppression
by impurities are very different from those occurring in diluted 
magnetic insulators, which are driven by classical percolation.\cite{Stinchcombe83} That is, 
if one is interested in singling out the geometrical 
aspects of impure superconductors, a model of correlated dilution should be
more appropriate.

\section{Finite-temperature properties: $T_c(f,U)$}
\label{finiteT}

Another consequence of the two-component nature of the order parameter is 
that at finite temperatures the superconducting-normal phase transition for the 
pure system belongs to the Kosterlitz-Thouless universality class. 
And as such, for $ 0 < T \leq T_c $, one expects that, asymptotically, 
\begin{equation}
\Gamma(r)\sim  r^{-\eta(T)},
\label{Gr2}
\end{equation}
where $\Gamma (r)$ is defined by Eq.\ (\ref{Gr}), and $\eta(T)$ increases monotonically between $\eta(0)=0$ and 
$\eta(T_c)=1/4$.\cite{Kosterlitz73,Berche02}
The finite-size scaling behavior of $P_s$ is therefore obtained upon 
integration of $\Gamma(r)$ over a two-dimensional system of linear 
dimension $L$. One then has \cite{Moreo91}
\begin{equation}
P_s=L^{2 - \eta(T_c)} F(L/ \xi), \ \ \ L\gg 1,\ T\to T_c^+, 
\label{Ps-fss}
\end{equation}
where $F(z)$ is a finite-size scaling function of the variable $z\equiv L/ \xi$, with 
\begin{equation}
\xi \sim \exp \left[ \frac{A}{(T-T_c)^{1/2}} \right],
\label{xi}
\end{equation}
where $A$ is a constant; in the thermodynamic limit, one recovers $P_s\sim\xi^{7/4}$. 
As discussed in Ref.\ \onlinecite{Paiva04}, we can obtain estimates of $T_c$ 
by plotting $L^{-7/4}P_s(L,\beta)$ as functions of $\beta$, for different $L$,
and by looking for intersections/merges of curves for consecutive values of $L$. This procedure was supported by independent estimates of the critical temperature through calculations of the superfluid stiffness, $D_s$, and using the universal jump at $T_c$; see Ref.\ \onlinecite{Paiva04} for details.

The general aspects of the universality class of the superconducting transition should
remain valid in the presence of disorder, since one still deals with a two-component order parameter. Further, numerical evidence has been gathered for the $xy$-model,\cite{Berche03} showing that $\eta(T_c)=1/4$ even in the presence of disorder; this is in agreement with the Harris criterion, which essentially states that disorder is irrelevant (in the renormalization group sense) if the specific heat exponent, $\alpha$, is positive.\cite{Harris74} 
In view of this, our data analyses for the finite-temperature transitions can follow along the same lines as those for the pure system,\cite{Paiva04} with both $P_s$ and $D_s$ now being understood 
as the configurationally-averaged equal-time pair correlation function and superfluid stiffness, respectively. 

\begin{figure}
{\centering\resizebox*{8.5cm}{!}{\includegraphics*{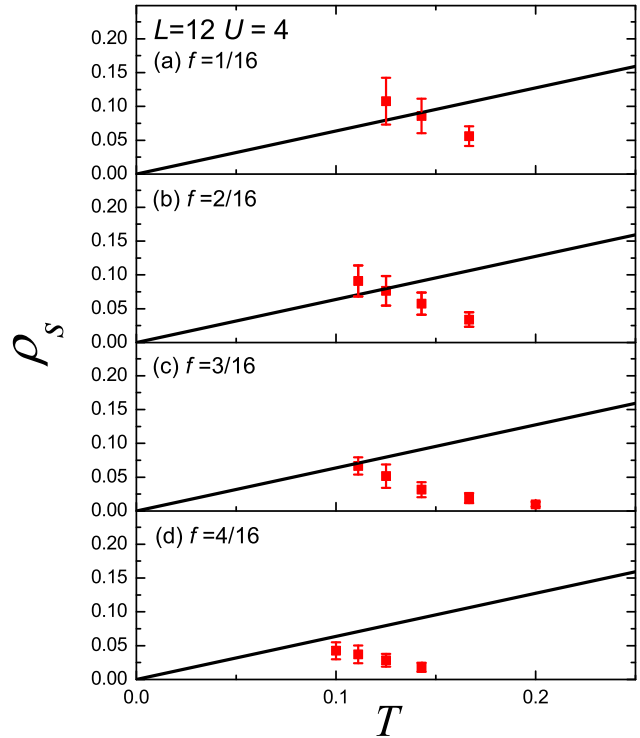}}}
\caption{(Color online) Configurationally-averaged helicity modulus as a function of temperature, for $U=4$ and different impurity concentrations $f$, for a $12\times 12$ lattice. 
In each panel, the straight line corresponds to $2T/\pi$.}
\label{rho_T} 
\end{figure}

\begin{figure}
{\centering\resizebox*{8.8cm}{!}{\includegraphics*{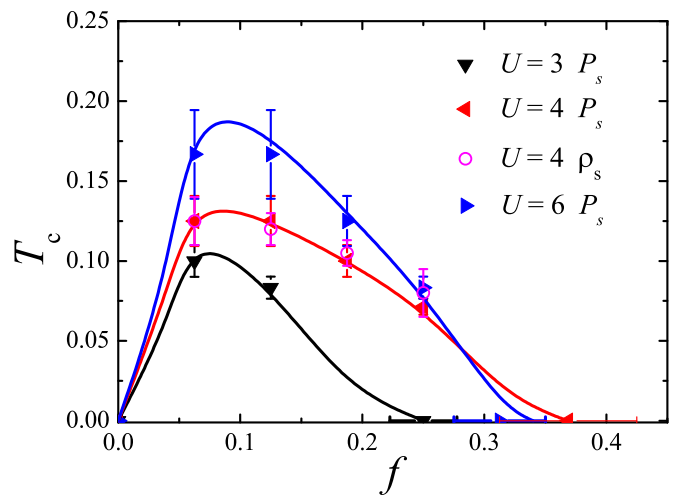}}}
\caption{(Color online) Critical temperature (in units of the bandwidth) for superconductivity, as a function of impurity concentration, $f$, for different values of the on-site attraction $U$. Full symbols have been obtained through the scaling of the pairing structure factor ($P_s$), whereas the empty circles correspond to data obtained through the helicity modulus ($\rho_s$). Full lines are guides to the eye.}
\label{Tcf} 
\end{figure}

Let us first consider the helicity modulus (HM),\cite{Scalapino93} which is given by 
\begin{equation}
\rho_s=\frac{D_s}{4\pi e^2},
\label{Ds}
\end{equation}
where $D_s$ is defined in Eq.\ (\ref{stiff}), and we take $e=1$ in our units. 
At the KT transition, the following universal-jump relation involving the 
helicity modulus holds: \cite{Nelson77}
\begin{equation}
T_c = \frac {\pi} {2} \rho_s^-,
\end{equation}
where $\rho_s^-$ is the value of the helicity modulus just below the
critical temperature. 
Thus, on universality grounds we may assume the same holds for a configurationally averaged $\rho_s$, and we can obtain $T_c$ by plotting $\rho_s(T)$, and looking for the intercept with $2T/\pi$. 
Figure \ref{rho_T} shows data for $U=4$, and we see that the intercepts occur at $T_c=0.125 \pm 0.015$, $0.12\pm 0.01$, $0.105 \pm 0.008$, and  $0.080 \pm 0.015$, for $f=1/16$, 2/16, 3/16 and 4/16, respectively. 
Similarly to the pure case,\cite{Paiva04} we have found here that finite-size effects are not too drastic, leading to essentially the same estimates for $T_c$. These estimates appear as empty circles in Fig.\ \ref{Tcf}.

\begin{figure}
{\centering\resizebox*{8.5cm}{!}{\includegraphics*{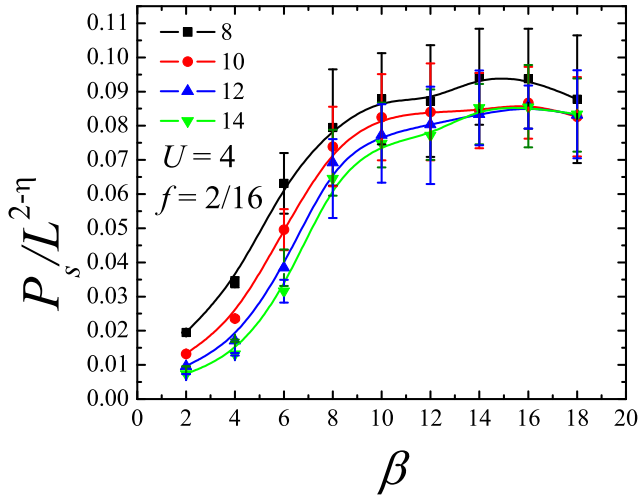}}}
\caption{(Color online) Scaled configurationally-averaged equal-time pair structure factor as a function of the inverse temperature $\beta$, for different lattice sizes ($L\times L$), at half filling, concentration of free sites $f=2/16$, and $U=4$. The curves are labelled by the linear lattice size $L$, and full lines are guides to the eye. }
\label{PseU4} 
\end{figure}

In Fig.\ \ref{PseU4} we show the scaled configurationally-averaged equal-time pair structure factor as a function of the inverse temperature $\beta$, for $U=4$ and different system sizes, for a given concentration of disorder. 
For usual second-order phase transitions, similar curves for two successive linear lattice sizes should cross at a single point, thus leading to estimates for critical inverse temperatures. 
For Kosterlitz-Thouless (KT) transitions, on the other hand, curves for different (but sufficiently large) lattice sizes should merge above a certain $\beta_c$.\cite{Barber83}
In the present case, we estimate $\beta_c$ as the smallest value for which the curves for the smallest size superimpose, within error bars, with the one for the largest size; this ensures that the error bars for data corresponding to intermediate sizes will also superimpose.
Thus, applying this criterion to the data in Fig.\ \ref{PseU4} yields $\beta_c = 8\pm 1$; 
this procedure is systematically repeated for other values of $f$, and we obtain the $T_c(f)$ data for $U=4$ shown in Fig.\ \ref{Tcf}. 
The estimates thus obtained are in excellent agreement with those obtained from the HM, thus adding credence to our merging criterion. 

\begin{figure}[h]
{\centering\resizebox*{8.5cm}{!}{\includegraphics*{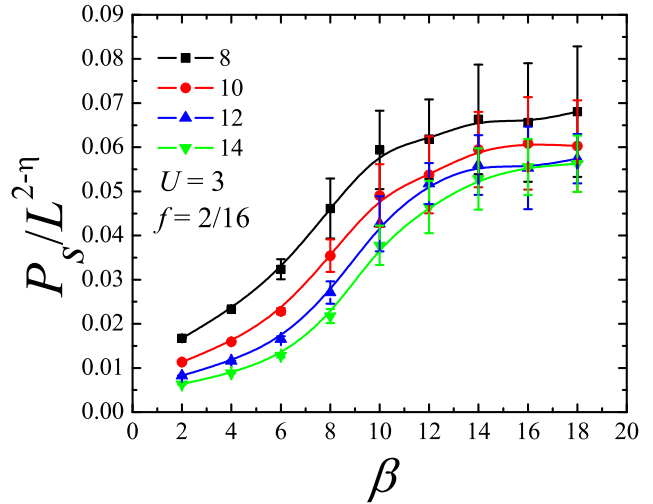}}}
\caption{(Color online) Same as Fig.\ \ref{PseU4}, but for $U=3$.}
\label{PseU3} 
\end{figure}

\begin{figure}
{\centering\resizebox*{8.5cm}{!}{\includegraphics*{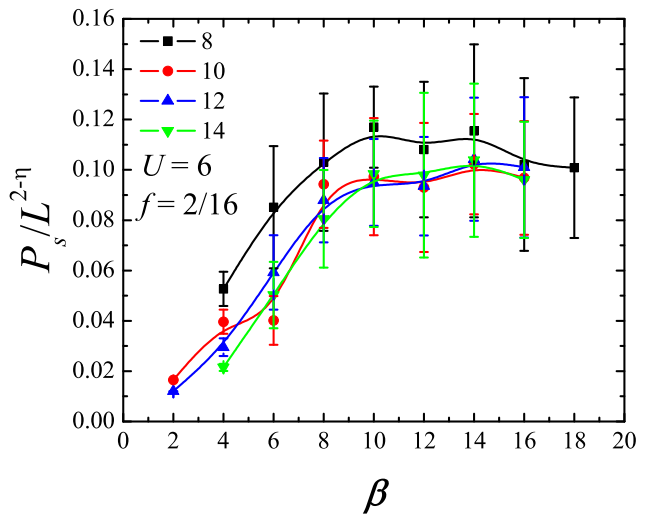}}}
\caption{(Color online) Same as Fig.\ \ref{PseU4}, but for $U=6$.}
\label{PseU6} 
\end{figure}

Given the fact that the calculations of configurationally averaged helicity moduli are very consuming in terms of computer time (for a given disorder configuration the CPU time is increased significantly due to the $\tau$-integration, and one performs averages over typically 50 disorder configurations), for other values of $U$ we only use data for $P_s$ to estimate $T_c(U,f)$. 
From Figs.\ \ref{PseU3} and \ref{PseU6} (which yield $\beta_c = 12\pm 1$ and $\beta_c = 6\pm 1$, respectively), as well as from similar ones for other values of $f$, we obtain the estimates for $T_c(f)$ for $U=3$ and $U=6$ shown in Fig.\ \ref{Tcf}.

Several interesting physical features emerge from Fig.\ \ref{Tcf}. 
Firstly, $T_c(0)=0$ as a result of the CDW-superconductivity degeneracy at half filling; however, any finite amount of disorder breaks this degeneracy and $T_c$ rises. 
Secondly, in all curves, $T_c$ displays a maximum at some $f_{\rm max}$, as a result of the interplay between the above-mentioned degeneracy and the behavior of the smallest energy scale, $\Delta$. 
Thirdly, we expect $T_c\to 0$ at $f_c(U)$, since above $f_c$, superconductivity cannot be sustained even in the ground state.
And, finally, near $f_c$, $T_c$ displays the convex shape observed in experiments;\cite{Haviland89} the steepness of the decrease in $T_c(f)$ can therefore be used to fit an effective $U$ by experimental data. 
As a final comment, one should have in mind that some of these results should change drastically as the system is doped away from half filling. Indeed, since in this case CDW-superconductivity degeneracy is already broken in the pure system, $T_c$ should display a monotonic decrease with $f$ for a given $U$; nonetheless, we can still expect $T_c$ to be a convex function of $f$.

\section{Conclusions}
\label{concs}

We have addressed the issue of disorder in two-dimensional superconductors. 
To this end, we have considered a simple model, namely the attractive Hubbard model, in which the on-site attraction is switched off on a fraction $f$ of sites, while keeping a finite $U$ on the remaining ones; the model is defined in such a way that $U>0$ in the attractive case [see Eq.\ (\ref{Ham})].
Through Quantum Monte Carlo simulations for typically 50 disorder configurations, we have calculated the configurational averages of the equal-time pair structure factor and, for $U=4$, the helicity modulus, as functions of temperature; there is no minus-sign problem in the attractive case. 
The continuous $O(2)$ symmetry of the superconducting order parameter allows us to use a spin-wave--like finite-size scaling form for the ground state behavior, from which the zero-temperature gap was calculated; at finite temperatures, the usual finite-size scaling form for the Kosterlitz-Thouless transition was used to calculate the critical temperature, which was checked for consistency against data for the helicity modulus.
Our numerical data are consistent with the following findings: 
(i) Superconductivity in the ground state is destroyed above an impurity concentration, $f_c$; 
(ii) At least up to $U=4$, this critical concentration increases with increasing $U$, slowly for $U\lesssim 2.5$, and then fast up to $U\sim 4$; this behavior does not agree with mean-field predictions, due to the important role played by fluctuations, not included in the latter approach. 
The error bars prevent us from ascertaining that $f_c$ decreases with $U$ above $U=4$, but it may be that the mean field behavior is recovered in this regime, since fluctuations should become less important for large $U$. 
At any rate, the transition at zero temperature is not driven by purely geometrical aspects, such as in dilute insulating magnets. 
(iii) In the range between $U=2.5$ and $U\lesssim6$, the normalized zero-temperature gap initially (i.e., small disorder) increases with disorder, as a result of both the breakdown of CDW-superconductivity degeneracy and the fact that free sites ``push" the electrons towards attractive sites;   
and, 
(iv) near the critical concentration of defects beyond its maximum value, $T_c$ is a convex function of $f$, as observed in experiments;  

Overall, we conclude that the random attractive Hubbard model is a promising working ground to investigate the interplay between impurities and pairing. By tuning two variables at half filling, namely the impurity concentration and the pairing potential, we have found instances in which small disorder either hardly affects superconductivity or enhances it. It should therefore be of interest to check whether these features remain valid away from half filling. Further, the present model can be used, with suitable changes, to investigate other disordered BCS superconductors, such as three-dimensional carbon-substituted MgB$_2$, \cite{Kazakov05} and MgB$_2$/MgO superstructures.\cite{Siemons08} 

\begin{acknowledgments}
This work was supported by 
the Brazilian Agencies CNPq, CAPES, FAPERJ, Instituto de 
Nanotecnologia/MCT, and Funda\c c\~ao Universit\'aria Jos\'e 
Bonif\'acio/UFRJ, and by 
DOE DE-FG01-06NA26204 and NSF OISE 0803230.
\end{acknowledgments}

\bibliography{biblio-sces}

\end{document}